%% file: main.tex
\documentclass[letterpaper]{article} 
\usepackage{aaai2027}  
\usepackage[hyphens]{url}  
\usepackage{graphicx} 
\usepackage{natbib}  
\usepackage{caption} 
\usepackage{algorithm}
\usepackage{algorithmic}
\usepackage{comment}
\usepackage{amsmath}
\usepackage{newfloat}
\usepackage{listings}
\DeclareCaptionStyle{ruled}{labelfont=normalfont,labelsep=colon,strut=off} 
\floatstyle{ruled}
\newfloat{listing}{tb}{lst}{}
\floatname{listing}{Listing}

\usepackage{booktabs}

\usepackage{xspace}
\newcommand{\tool}{{\it Ledger}\xspace}

\usepackage{xcolor}

\title{Turning Interaction History into Execution State: A Runtime Layer for Long-Horizon Coding Agents}
\author{
    Zehao Wang\textsuperscript{\rm 1},
    Yisen Xu\textsuperscript{\rm 1},
    Chenglin Li\textsuperscript{\rm 1},
    Chao Peng\textsuperscript{\rm 2},
    Bram Adams\textsuperscript{\rm 3},
    Ahmed E. Hassan\textsuperscript{\rm 3},
    Tse-Hsun (Peter) Chen\textsuperscript{\rm 1}
}
\affiliations{
    \textsuperscript{\rm 1}Concordia University, Montreal, Quebec, Canada\\
    \textsuperscript{\rm 2}Tencent, China\\
    \textsuperscript{\rm 3}School of Computing, Queen's University, Kingston, ON, Canada\\
    w\_zeha@encs.concordia.ca, yisen.xu@mail.concordia.ca, chenglin.li@mail.concordia.ca,\\
    chao.peng@acm.org, bram.adams@queensu.ca, ahmed@cs.queensu.ca, peterc@encs.concordia.ca
}

\begin{document}
\nocopyright
\maketitle

\begin{abstract}
Long-horizon coding agents accumulate hundreds of actions and observations in their trajectories, yet nothing in this record indicates which observations still describe the repository as it currently stands. Before every decision, the model must implicitly infer the execution status from raw history, and when this inference falls short, the agent acts on outdated file contents or re-executes work whose results are still valid. We propose \tool, a deterministic runtime layer that distills an agent's completed interactions into an explicit \emph{execution state}: what has been observed, what has been modified, and what has been attempted. \tool keeps this state in an online execution ledger and applies it at two boundaries of every step. Before the model acts, an \emph{inform} path appends a compact runtime state view to the prompt; before a proposed command runs, a \emph{govern} path checks it against the ledger, returning still-valid earlier results in place of re-execution and flagging likely-redundant repetition. The layer adds no language-model calls and wraps an otherwise unmodified agent. Across all 500 SWE-bench Verified instances, \tool raises Pass@1 from 56.2\% to 64.2\% with GPT-5 mini and from 75.8\% to 81.0\% with MiniMax M2.5, while cutting total cost by 28.9\% and 31.8\%. Attached to OpenAI Codex, it adds 3.4 percentage points of Pass@1 at 24.4\% lower cost. Ablations attribute most of the resolution gain to \emph{govern} and most of the efficiency gain to \emph{inform}, with their combination performing best. What long-horizon agents lack, we conclude, is not a shorter view of their history but an explicit account of their own execution state.
\end{abstract}

\input{intro}

\input{related}

\input{method}

\input{evaluation}

\input{limitation}
\input{conclusion}

\bibliography{aaai2027}


\end{document}

%% file: intro.tex
\section{Introduction}

Large language models now power agents that resolve issues in real software
repositories~\citep{yang2024sweagent,yao2023react,wang2025openhands,zhang2024autocoderover}.
These agents construct repairs through extended interactions with the
repository. Their trajectories preserve the sequence of actions and
observations, but not which parts of that history still describe the current
execution. As the interaction progresses, later actions can change whether
earlier observations remain current, while prior commands and their results
become dispersed across an increasingly long trajectory. The model must
therefore reconstruct where the execution stands before deciding what to do
next. When this reconstruction is incomplete, the agent may rely on stale
observations or repeat work whose result remains valid.

Existing approaches make long trajectories easier to process by truncating or
summarizing the interaction history presented to
the model~\citep{lian2026swe,wang2026swe,xiao2026reducing}. These methods
reduce how much history the model must reason over, but they do not maintain
an explicit account of the execution as it evolves. Changing the history
presented to the model therefore does not determine which prior observations
remain current or whether a proposed action repeats work already completed
under the same conditions. Addressing this limitation requires maintaining
execution state across steps and applying it both when the model chooses its
next action and before that action executes.

We present \textbf{\tool}, a
runtime layer that turns the agent's actions and observations into an
\emph{execution state} capturing what it has observed, what has changed, and
what it has already attempted. \tool maintains this state in a deterministic
\emph{execution ledger} and applies it at two points in each agent step.
Before action generation, an \emph{inform} path renders a compact state view
into the model input, making prior observations and their current status
directly available to the agent. Before execution, a \emph{govern} path
evaluates the proposed command against the same ledger, conservatively reuses
an earlier result when it remains valid, and provides guidance when a
repetition may be unnecessary. The language model remains responsible for
understanding the issue and constructing the repair, while the runtime
maintains only facts that can be derived mechanically from completed
interactions. \tool requires no additional language-model calls and can be
integrated with an agent scaffold without modifying its internal logic and workflow.

We evaluate \tool on all 500 instances of SWE-bench Verified~\citep{jimenez2024swe}. Our results show that \textbf{(1)} on mini-swe-agent, \tool increases Pass@1 from 56.2\% to 64.2\% with GPT-5 mini and from 75.8\% to 81.0\% with MiniMax M2.5, corresponding to relative reductions in unresolved instances of 18.3\% and 21.5\%; \textbf{(2)} total cost decreases by 28.9\% and 31.8\%, respectively, showing that the resolution gains do not come at the expense of higher execution cost; \textbf{(3)} integrating the same execution-state layer with OpenAI Codex increases Pass@1 by 3.4 percentage points while reducing total cost by 24.4\%, demonstrating that the gains extend beyond a single agent scaffold; and \textbf{(4)} an ablation shows that \emph{govern} drives most of the resolution gains, \emph{inform} primarily improves execution efficiency, and their combination provides the strongest overall performance.

Across two backbones and two substantially different agent scaffolds, \tool improves repair success while reducing execution cost.

We make the following contributions:
\begin{itemize}
\item We introduce \emph{execution state} as a representation of where a
coding agent currently stands, derived deterministically from its prior
interactions. Unlike the trajectory itself, this state reflects how earlier
observations and actions apply at the current step.

\item We propose \tool, a deterministic runtime layer that maintains this
state in an online execution ledger and applies it through complementary
\emph{inform} and \emph{govern} paths. \tool requires no additional
language-model calls and integrates with an otherwise unmodified agent.

\item We evaluate \tool on all 500 SWE-bench Verified instances using two
backbones and two agent scaffolds. \tool improves Pass@1 by up to 8.0
percentage points while reducing total cost by up to 31.8\%, with consistent
gains on both mini-swe-agent and OpenAI Codex.

\end{itemize}

%% file: related.tex
\section{Related Work}
\label{sec:related}

\subsection{LLM Agents for Software Issue Resolution}
A rapidly growing line of work builds LLM agents that resolve real-world repository issues, differing mainly in where they place the intelligence. Some refine the agent-computer interface, giving the model dedicated commands for search, navigation, and editing \citep{yang2024sweagent} or unifying its action space as executable code \citep{wang2024executable, wang2025openhands}. Others strengthen localization and repair, through program-structure-aware search \citep{zhang2024autocoderover}, repository knowledge graphs \citep{ma2025alibaba}, or autonomous planning of repair steps \citep{bouzenia2025repairagent}. A third group replaces the interactive loop altogether with fixed pipelines \citep{xia2024agentless} or ensemble generation and selection \citep{gao2025trae}. All of these efforts improve the agent itself. \tool sits at a different level: it is a runtime layer around an unmodified agent, so it neither competes with nor requires changes to any of these designs, and our transfer experiment shows the same layer attaching to two different scaffolds.

\subsection{Behavior of Long-Horizon Coding Agents}

Recent studies characterize how coding agents behave over long runs. \citet{bouzenia2025understanding} analyze 120 trajectories and 2{,}822 model interactions across three agents and report recurring action sequences, meaning that the same operations reappear within a single run. \citet{majgaonkar2025understanding} find that failed trajectories are consistently longer and more variable than successful ones, even though the file that needs to change is already identified in 72\% to 81\% of the runs that ultimately fail. The additional length in those runs is therefore not spent locating the defect. It reflects effort spent again on ground the agent has already covered. These studies establish the behavior we target, but they remain diagnostic. The agent itself is still given no way to tell that a file it is about to open is one it has already read, and \tool closes exactly this gap.

\subsection{Context Management for LLM Agents}

A growing body of work reduces the text an agent reads at each step. Below the agent level, prompt- and KV-level compression shortens what the model attends to without awareness of the agent's interaction structure \citep{jiang2023llmlingua, xiao2024efficient}. The methods we discuss next instead operate on the agent trajectory itself. SWE-AGILE keeps a sliding window of full reasoning and compresses older reasoning into short digests, which limits context growth for reasoning models without forcing the agent to derive earlier conclusions again \citep{lian2026swe}. SWE-Pruner trains a small dedicated model that selects the task-relevant lines of a long context and reports token reductions of roughly a quarter to a half on multi-turn tasks \citep{wang2026swe}. AgentDiet removes trajectory content that it identifies as redundant or expired, cutting input tokens by roughly 40\% to 60\% at comparable task performance \citep{xiao2026reducing}. Against this trend toward more elaborate schemes, \citet{lindenbauer2025complexity} report a cautionary result. Simply masking old observations achieves the same solve rate as LLM-generated summarization while roughly halving the cost of the raw agent. This calls into question whether learned or model-generated context compression is worth its overhead. 

These methods act on what the agent reads and leave what the agent does untouched. Compressing a transcript makes a repeated read cheaper, but it does not prevent the read. Several of them also decide what to keep with a learned or model-generated estimate of relevance, which adds inference cost of its own. \tool takes a complementary approach. It screens each command before execution and prevents redundant commands from running, and it bases this decision on a record of what the agent has actually read and changed rather than an estimate of what looks relevant. This record does not require any model calls to maintain, in line with the evidence above that simple deterministic strategies compete well with model-generated compression.

%% file: method.tex
\section{\tool: Execution-State Management for Coding Agents}
\label{sec:method}

A coding agent accumulates reads, edits, tests, and other commands throughout execution, with the resulting actions and observations accumulated in the model context. This context preserves what happened but does not explicitly maintain the current implications of those interactions: whether previously observed code remains valid after later edits, whether an earlier result can still be reused, or whether the next command would repeat previously completed work. As the context grows, these execution facts become distributed across the trajectory, increasing the risk that the model overlooks or misapplies the information that is no longer valid. 

We present \textbf{\tool}, a deterministic execution-state layer for long-horizon coding agents. Here, \emph{execution state} denotes mechanically derivable facts about the agent's prior observations, repository modifications,
and attempted actions, rather than the model's latent reasoning state or a complete representation of the environment. As the agent interacts with the repository, \tool incrementally maintains an \emph{execution ledger} that records what it has observed, what has changed, and what it has already attempted, together with the current status of those interactions. The \emph{inform} path renders a compact view of this state at the end of the model context before action generation, while the \emph{govern} path evaluates each proposed command against the same state before execution and mediates redundant or potentially unproductive repetition.

\begin{figure*}[t]
  \centering
  \includegraphics[width=0.7\textwidth]{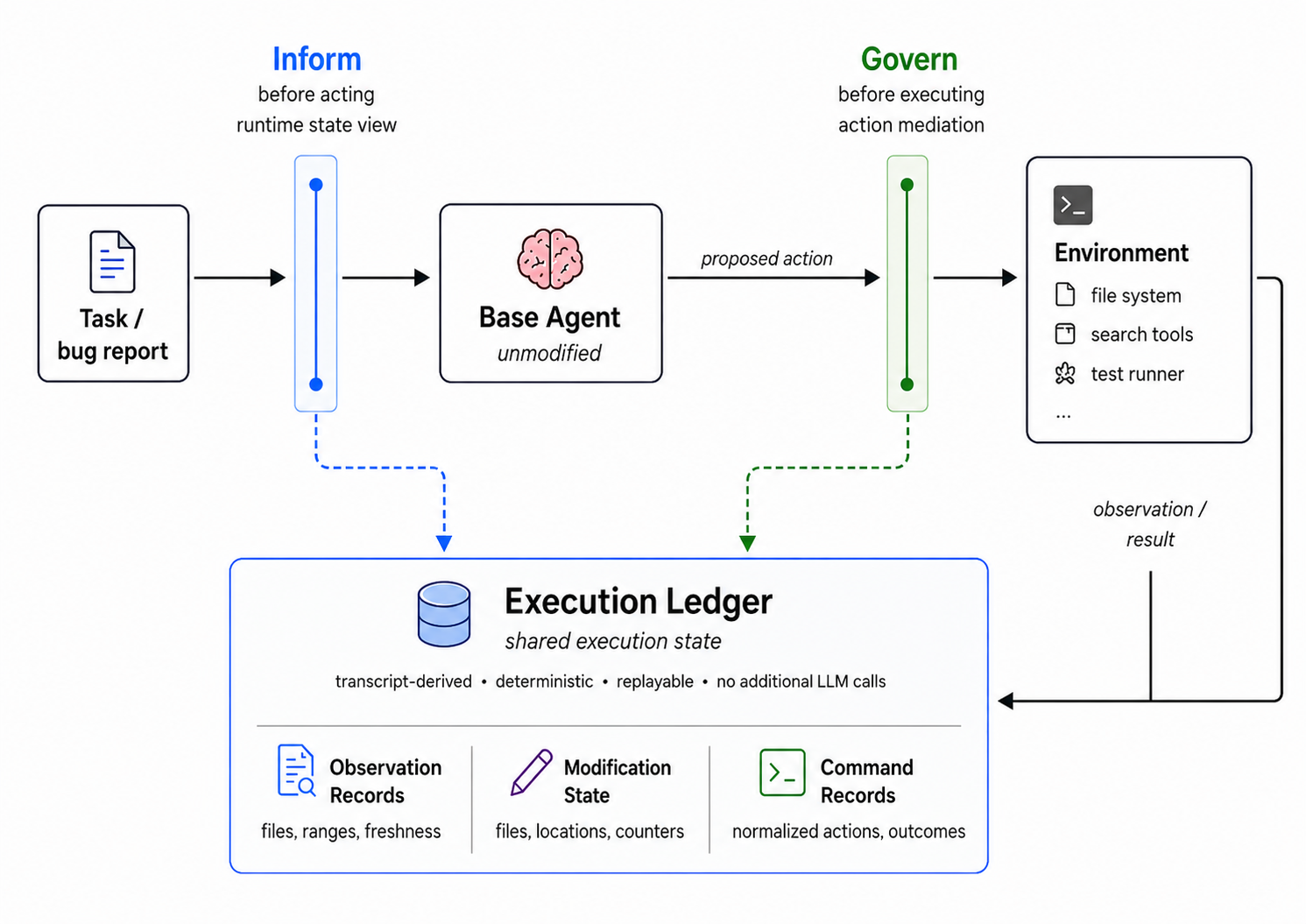}
  \caption{Overview of \tool. A single execution ledger records the agent's
  read, edit, and command events. It is the only state \tool adds, and it
  feeds two paths: an \emph{inform} path that renders a runtime state view into the
  model input, and a \emph{govern} path that evaluates each proposed command
  before it executes. Both paths are computed from the ledger and the
  transcript. The base agent is unmodified and remains the only language
  model in the loop.}
  \label{fig:overview}
\end{figure*}

\tool separates semantic repair reasoning from deterministic execution-state management. The language model remains responsible for understanding the task and constructing the repair, while \tool maintains the mechanically derivable execution state induced by prior interactions. It requires no modification to the base agent, introduces no additional language-model calls, and applies across models and agent scaffolds.

\subsection{Online Execution-State Management}
\label{sec:runtime}

We consider a coding agent that interacts with a repository through a sequence of tool calls. Before step $t$, its persistent interaction history is
\[
H_t=(m_0,a_0,o_0,\ldots,a_{t-1},o_{t-1}),
\]
where $m_0$ is the task description and each pair $(a_i,o_i)$ contains an action proposed by the agent and the corresponding runtime observation. A standard agent selects its next action from this accumulated history.

\tool maintains an execution ledger $L_t$ online alongside $H_t$. Before action generation, the \emph{inform} path constructs a compact state view
\[
M_t=\operatorname{Render}(m_0,L_t),
\]
and places it after the persistent history in the current model input. The agent therefore selects its next action using both the accumulated interaction history and the current execution-state view.

Before the proposed action reaches the execution environment (the repository and shell), the \emph{govern} path evaluates it against the same ledger:
\[
d_t=\operatorname{Govern}(a_t,L_t).
\]
The runtime then applies the decision $d_t$ to mediate the proposed action, either executing it unchanged, executing it while augmenting the resulting observation, or suppressing it and returning a reference to an earlier result, as detailed in Section~\ref{sec:govern}. The resulting interaction (e.g., the action-observation pair) is appended to the persistent history, and the ledger is updated from the proposed action, governing decision, and returned observation. Thus, the same online execution state is used at both boundaries of each step, allowing the inform path to expose prior execution state before action generation and the govern path to mediate redundant actions that are still proposed.

\subsection{Online Execution Ledger}
\label{sec:ledger}

The execution ledger $L_t$ is an online state representation updated deterministically from completed runtime interactions. It comprises three complementary components: \emph{observation records}, which track the portions of each file actually returned to the agent rather than merely requested; \emph{modification state}, which tracks repository changes; and \emph{command records}, which represent actions previously attempted by the agent. These components are derived directly from actions and runtime observations, require no language-model calls, and can be reconstructed from a saved trajectory.

\paragraph{Observation records.}
An observation record identifies a file, the portion of that file returned to the agent, and the change-counter values at the time of observation (defined under \emph{Modification state} below). The observed portion is represented either as the complete file or as an explicit line range. A record is created only when a command successfully returns the contents of a single file without truncation; failed, ambiguous, and truncated reads are excluded. Observation records therefore capture verified coverage rather than merely recording that a read command was issued.

\paragraph{Modification state.}
The ledger tracks the files, and where detectable the symbols, modified by the agent. Each tracked file has a local change counter that advances whenever the file is edited, while a global counter advances after any tracked repository modification. Observation records retain both counter values from the time they were created. The local counter detects direct changes to the observed file, while the global counter provides a conservative indication that the repository has changed since the observation. Comparing the stored and current values allows the ledger to distinguish observations that remain current from those that may have been invalidated by subsequent changes. Transient artifacts, such as scratch files and backups, are excluded based on file names and extensions.

\paragraph{Command records.}
A command record contains a normalized representation of an issued command, its operational category, and its position in the recent execution history. Normalization removes incidental differences, such as a leading directory change, while categories distinguish operations such as file inspection, search, testing, and modification. Together with the modification state, command records distinguish actions repeated after repository changes from actions repeated under unchanged conditions. The distinction matters because a command repeated after an edit, such as rerunning a failing test after a fix, may provide new evidence, whereas the same command repeated without an intervening change typically reproduces its earlier result. They support the detection of exact repeats, repeated tests without intervening edits, and short command loops without relying solely on literal command-string equality.

Together, these components form the shared execution state used by the inform and govern paths.

\subsection{Inform: Runtime State View}
\label{sec:inform}

The inform path exposes the current execution state to the language model before action generation. At step $t$, \tool renders the state view $M_t$ from the task description $m_0$ and ledger $L_t$, then places it at the end of the current model input. The view is regenerated rather than appended to the persistent trajectory because its contents can change over time: an observation that is current at one step may become stale after a later edit. Previous views therefore do not accumulate, and each ledger update is reflected in the next rendering. Because the view occupies only the end of the model input, re-rendering it leaves the append-only prefix unchanged and therefore does not disturb prompt caching.

The view contains a fixed task anchor and two ledger-derived components. The \emph{current focus} lists the files most recently modified by the agent. The \emph{observation index} records which files or line ranges have already been observed and reports the current status of those observations.

For each observation, the view distinguishes content that remains current from content that may have been invalidated by subsequent changes. A complete read identifies the file as fully observed, while a partial read retains the covered line range. When neither the stored file counter nor the stored global counter differs from its current value, the observation is marked as current. When subsequent changes prevent this guarantee, the view indicates that the earlier observation may be stale. If the ledger cannot determine the status reliably, it makes no freshness claim.

By consolidating these execution facts at a consistent position in the context, the inform path makes prior observations and their current status directly available during action generation. The view remains advisory and does not constrain the agent's action space; the govern path provides runtime intervention when guidance alone does not prevent redundant work.

\subsection{Govern: State-Conditioned Action Mediation}
\label{sec:govern}

The govern path evaluates each proposed command before it reaches the environment. Given the proposed action $a_t$ and current ledger $L_t$, it produces a runtime decision

\[
d_t=\operatorname{Govern}(a_t,L_t)
\]

from three possible outcomes. \textsc{Allow} executes the command unchanged. \textsc{Reuse} suppresses the command and returns a reference to a still-visible result from an earlier equivalent interaction. \textsc{Nudge} allows the command to execute but augments the resulting observation with a short message indicating that the action may repeat prior work.

\paragraph{Previously observed content.}
When the agent proposes reading content already covered by an observation record, govern checks whether the corresponding observation remains current. A complete read may receive \textsc{Reuse} when the proposed file has not changed and the earlier output remains visible in the model context. A partial read is reusable only when the proposed line range is fully covered by an earlier observation. Otherwise, the command is allowed. Because \textsc{Reuse} returns only a reference to the earlier output, govern never points the agent to content that is no longer present in its context.

\paragraph{Repeated commands.}
Govern uses command records and modification state to identify actions repeated under unchanged execution conditions. 
\textsc{Reuse} applies only to inspection and search commands, as classified by their command records (Section~3.2). These commands do not modify the repository, so under an unchanged modification state an exact repeat reproduces its earlier output, and the recorded result can be returned in place of execution. A test command repeated without an intervening edit receives \textsc{Nudge} instead,
because rerunning the test may still be intentional despite the unchanged code state. Repeated normalized commands within a short execution window likewise receive \textsc{Nudge} as a possible execution loop. Thus, the policy suppresses only repetitions whose prior results can be reused safely, while potentially intentional repetitions remain executable.

\paragraph{Conservative safeguards.}
The policy is designed to avoid obstructing valid exploration. Commands used to configure the environment, reproduce the reported problem, generate or submit a repair, or follow newly observed errors are always allowed. Interventions are capped per task, and repeated nudges are subject to a cooldown. If the agent repeats a read after receiving a reference to an earlier result, the next attempt is allowed, preventing cycles between the agent and the runtime. Unsupported commands, parsing failures, and internal policy errors default to \textsc{Allow}.

The govern path therefore applies two forms of mediation. It suppresses actions only when an earlier result can be reused conservatively, and it nudges actions when the ledger indicates a potentially unproductive repetition. It does not assess the semantic correctness of the repair strategy; it mediates only execution patterns supported by the recorded interaction state.

%% file: evaluation.tex
\begin{table}[t]
\centering
\small
\setlength{\tabcolsep}{4pt}
\begin{tabular}{llccc}
\toprule
Method & Backbone & Pass@1 & $\Delta$  & RRU \\
\midrule
\multicolumn{5}{l}{\emph{Results reported in prior work}} \\
\textsc{SWE-Pruner}
& Claude Sonnet 4.5 & 70.6 $\rightarrow$ 72.0 & $+1.4$ & 4.8\%\\
& GLM-4.6           & 55.4 $\rightarrow$ 56.6 & $+1.2$ & 2.7\%\\
\textsc{LaMR}
& Claude Sonnet 4.5 & 70.6 $\rightarrow$ 71.8 & $+1.2$ & 4.1\% \\
& Claude Opus 4.6   & 75.6 $\rightarrow$ 76.0 & $+0.4$ & 1.6\%\\
\midrule
\multicolumn{5}{l}{\emph{Our results}} \\
\tool
& GPT-5 mini        & 56.2 $\rightarrow$ 64.2 & $+8.0$ & 18.3\%\\
& MiniMax M2.5      & 75.8  $\rightarrow$ 81.0 & $+5.2$ & 21.5\%\\
\bottomrule
\end{tabular}
\caption{Reported Pass@1 on SWE-bench Verified, shown as base $\rightarrow$ method, with $\Delta$ denoting the percentage-point improvement. RRU shows the relative reduction in unresolved instances. Results for prior methods are taken from their original papers; all use mini-swe-agent and the full 500-instance benchmark but differ in their backbones and evaluation setups.}
\label{tab:main}
\end{table}

\begin{table*}[t]
\centering
\small
\begin{tabular}{lrrrrrr}
\toprule
 & \multicolumn{3}{c}{GPT-5 mini} & \multicolumn{3}{c}{MiniMax M2.5} \\
\cmidrule(lr){2-4} \cmidrule(lr){5-7}
Totals over 500 instances & base & $+$\tool & $\Delta$ & base & $+$\tool & $\Delta$ \\
\midrule
Total cost (USD)            & \$23.60  & \$16.78  & $-28.9\%$ & \$34.31  & \$23.39  & $-31.8\%$ \\
Cost per resolved instance  & \$0.084  & \$0.052  & $-37.8\%$ & \$0.091  & \$0.058  & $-36.2\%$ \\
Input tokens                & 131.7M   & 110.7M   & $-15.9\%$ & 616.7M   & 289.4M   & $-53.1\%$ \\
Model calls                 & 10{,}171 & 8{,}986  & $-11.7\%$ & 30{,}219 & 20{,}566 & $-31.9\%$ \\
Redundant snippet re-reads  & 1{,}875  & 1{,}421  & $-24.2\%$ & 5{,}840  & 3{,}790  & $-35.1\%$ \\
\bottomrule
\end{tabular}
\caption{Efficiency totals across all 500 instances on mini-swe-agent v2. All changes are relative percentages; comparisons within each backbone use identical pricing assumptions. \tool reduces cost on both backbones while improving resolution (Table~\ref{tab:main}).}
\label{tab:efficiency}
\end{table*}

\section{Experiments}
\label{sec:experiments}

\subsection{Experimental Setup}

\paragraph{Benchmark and agent scaffolds.}
We evaluate on all 500 instances of SWE-bench Verified~\citep{jimenez2024swe}. Our primary controlled evaluation uses mini-swe-agent v2~\citep{yang2024sweagent}, a minimal scaffold in which actions are issued as Bash commands. \tool is
integrated without changing the task prompt or the available tools. We additionally integrate \tool with the state-of-the-art OpenAI Codex, which represents actions as structured tool calls, to test whether the layer transfers to a substantially different agent scaffold.

\paragraph{Models and configurations.}
For mini-swe-agent, we evaluate GPT-5 mini and MiniMax M2.5 with high reasoning effort. For Codex, we evaluate MiniMax M2.5 as it is the stronger model of the two. In all comparisons, the prompt, tool, and configuration remain the same. 
All patches are graded with the official SWE-bench test
suites.

\paragraph{Metrics.}
Our primary metric is \textit{\textbf{Pass@1}}, the fraction of instances resolved by a single
agent run. Because stronger baselines leave fewer failures available to
recover, we also report the \textit{\textbf{relative reduction in unresolved instances (RRU)}}~\cite{manning1999foundations},
computed as
\[ \mathrm{RRU}=
\frac{\mathrm{Pass@1}_{\tool}-\mathrm{Pass@1}_{\mathrm{base}}}
{1-\mathrm{Pass@1}_{\mathrm{base}}}.
\]
We further report paired outcomes: the baseline failures recovered by
\tool and the baseline successes on which it regresses.

We measure efficiency using total API cost, input tokens, model calls, and executed commands over the full benchmark. We also report cost per resolved instance and redundant snippet re-reads, defined as requests for a previously returned file range that has not changed since it was last observed. Comparisons within each model and agent scaffold use identical pricing
assumptions.

\paragraph{Baselines.} To better position \tool among recent context-management approaches, we also include two methods previously evaluated on SWE-bench Verified. SWE-Pruner~\cite{wang2026swe} prunes file-reading outputs to lines relevant to the current goal. LaMR~\cite{lamr} filters such outputs while preserving the structural dependencies needed to interpret the retained code. Table~\ref{tab:main} reports the results published in the respective papers.

\subsection{Results on mini-swe-agent v2}
\paragraph{Resolution.}
\tool improves Pass@1 with both backbones (Table~\ref{tab:main}). With
GPT-5 mini, Pass@1 increases from 56.2\% to 64.2\%. \tool recovers 59 instances that the baseline fails and regresses on 19 that it solves, yielding a net gain of 40. With MiniMax M2.5, Pass@1 increases from 75.8\% to 81.0\%, with 38  recoveries and 12 regressions, yielding a net gain of 26. Both improvements are statistically significant under McNemar's test ($p<0.001$ for GPT-5 mini and $p<0.01$ for MiniMax M2.5).

Because MiniMax M2.5 starts from a stronger baseline and leaves fewer
instances unresolved, we also report the relative reduction in unresolved
instances. \tool eliminates 18.3\% of the failures remaining under GPT-5 mini
and 21.5\% of those remaining under MiniMax M2.5. These reductions are
similar across the two backbones despite their different baseline resolution
rates.

\paragraph{Comparison with context-management methods.}
Table~\ref{tab:main} places our results alongside the gains reported by recent context-management methods on SWE-bench Verified. Because these studies use different backbones, agent scaffolds, and evaluation settings, the numbers are not directly comparable. Nevertheless, these methods report improvements of $0.4$--$1.4$ percentage points over their respective base agents, whereas \tool improves its base scaffold by $5.2$--$8.0$ points. The RRU of \tool is also much larger than prior methods (18.3\%--21.5\% vs. 1.6\%--4.8\%). 

The methods also intervene at a different point in the agent loop. Prior work changes what the agent reads by compressing observations or trimming the interaction history. In contrast, \tool tracks the agent's execution state, including which files have been viewed and which edits have been applied, and uses this state to prevent redundant actions.

Because \tool adds an execution layer to the base agent, we examine whether its resolution gains come at additional inference cost. As shown in Table~\ref{tab:efficiency}, \tool improves resolution while reducing both total cost and model usage. Total cost decreases by 28.9\% with GPT-5 mini and 31.8\% with MiniMax M2.5, while model calls decrease by 11.7\% and 31.9\%, respectively. Since more instances are resolved at the same time, the cost per resolved instance falls by 37.8\% and 36.2\%. The reduction is especially large with MiniMax M2.5, where \tool uses 53.1\% fewer input tokens while resolving 26 additional instances.

These savings are partly explained by fewer repeated inspections. The GPT-5 mini and MiniMax M2.5 baselines perform 1{,}875 and 5{,}840 redundant snippet re-reads, respectively; \tool lowers these totals by 24.2\% and 35.1\%. This shows that repeated inspection is common under both backbones and can be reduced without sacrificing resolution. Although \tool is designed primarily to improve execution reliability by keeping the agent consistent with what has already occurred, lower token usage and cost emerge as a secondary benefit of avoiding redundant actions.

\subsection{Results on OpenAI Codex}
\label{sec:codex}

To test whether \tool depends on mini-swe-agent, we integrate it with OpenAI
Codex, which uses a substantially different agent design. The execution
ledger, governing policy, and state-view renderer remain unchanged; only a
Codex-specific adapter and a lightweight model-API proxy are added. We
evaluate both configurations on all 500 instances using MiniMax M2.5, the
stronger of the two backbones.

\paragraph{Resolution.}
\tool increases Codex Pass@1 from 74.8\% to 78.2\%, a gain of 3.4
percentage points that eliminates 13.5\% of the failures remaining under the
baseline (Table~\ref{tab:codex}). The improvement is statistically significant under McNemar's test ($p<0.05$). This shows that the same execution-state layer improves resolution in a  different coding agent without changing its core components.

\paragraph{Efficiency.}
\tool reduces total cost by 24.4\%, input tokens by 33.1\%, model calls by
17.9\%, and redundant re-reads by 6.1\%, lowering cost per resolved instance
by 27.7\%. The efficiency gains are smaller than with mini-swe-agent on the
same backbone, particularly for redundant re-reads, which fall by 6.1\%
rather than 35.1\%. Cost and input tokens nevertheless decrease
substantially, indicating that the Codex saving comes more from shortening
executions overall than from eliminating repeated inspection specifically.
Across both agents, \tool consistently improves resolution while lowering
execution cost.

\begin{table}[t]
\centering
\small
\setlength{\tabcolsep}{3pt}
\begin{tabular}{lrrr}
\toprule
Codex: totals over 500 & base & $+$\tool & $\Delta$ \\
\midrule
Resolved                   & 374 (74.8\%) & \textbf{391 (78.2\%)} & $+3.4$ pp \\
Unresolved (RRU)           & 126          & \textbf{109}          & $-13.5\%$ \\
Total cost (USD)           & \$30.18      & \textbf{\$22.82}      & $-24.4\%$ \\
Cost per resolved instance & \$0.081      & \textbf{\$0.058}      & $-27.7\%$ \\
Input tokens               & 636.5M       & \textbf{425.6M}       & $-33.1\%$ \\
Model calls                & 25{,}271     & \textbf{20{,}736}     & $-17.9\%$ \\
Redundant snippet re-reads        & 5{,}097      & \textbf{4{,}784}      & $-6.1\%$ \\
\bottomrule
\end{tabular}
\caption{Integrating \tool with OpenAI Codex on all 500 instances of
SWE-bench Verified using MiniMax M2.5.}
\label{tab:codex}
\end{table}

\begin{table}[t]
\centering
\small
\setlength{\tabcolsep}{1pt}
\begin{tabular}{lccccc}
\toprule
Method & Resolved & Pass@1 & Calls & In. tok. & Cost \\
\midrule
\multicolumn{6}{l}{\emph{GPT-5 mini}} \\
mini-swe-agent
  & 281/500 & 56.2 & 10.1K & 131.7M & \$23.60 \\
$+$ \tool\ (inform only)
  & 320/500 & 64.0 & \textbf{8.9K} & \textbf{106.2M} & \$16.09 \\
$+$ \tool\ (govern only)
  & 321/500 & 64.2 & 10.8K & 149.7M & \textbf{\$15.67} \\

$+$ \tool\ (inform+govern)
  & \textbf{321/500} & \textbf{64.2} & \textbf{8.9K}
  & 110.7M & \$16.78 \\
\midrule
\multicolumn{6}{l}{\emph{MiniMax M2.5}} \\
mini-swe-agent
  & 379/500 & 75.8 & 30.2K & 616.7M & \$34.31 \\
$+$ \tool\ (inform only)
  & 385/500 & 77.0 & 21.3K & 336.4M & \$25.71 \\
$+$ \tool\ (govern only)
  & 402/500 & 80.4 & 29.2K & 589.2M & \$30.80 \\
$+$ \tool\ (inform+govern)
  & \textbf{405/500} & \textbf{81.0} & \textbf{20.6K}
  & \textbf{289.4M} & \textbf{\$23.39} \\
\bottomrule
\end{tabular}
\caption{Ablation results over the same 500 SWE-bench Verified instances
using mini-swe-agent. Model calls, input tokens, and cost are reported as
totals across all 500 instances. \emph{Govern only} retains the governing policy and removes the runtime
state view; \emph{inform+govern} enables both paths and is the full system.}
\label{tab:ablation}
\end{table}

\subsection{Ablation}
\label{sec:ablation}

\tool combines two paths: \emph{inform}, which exposes the current runtime
state to the agent, and \emph{govern}, which mediates proposed redundant
actions. To examine their contributions, we evaluate an \emph{inform-only}
configuration that retains the state view but removes the governing policy,
and a \emph{govern-only} configuration that retains the governing policy but
removes the state view. We run both ablations on mini-swe-agent with both
backbones over the same 500 SWE-bench Verified instances under the same
protocol (Table~\ref{tab:ablation}).

\paragraph{Govern contributes the larger and more consistent resolution gain.}
With GPT-5 mini, inform-only resolves 320 instances, while govern-only and the
full system each resolve 321, compared with 281 for the baseline. Thus, either
path independently captures nearly the entire gain on this backbone. With
MiniMax M2.5, inform-only resolves 385 instances and govern-only resolves 402,
compared with 405 for the full system and 379 for the baseline. Govern
therefore captures most of the MiniMax M2.5 gain, while adding the state view
contributes three further resolutions. These results show that both paths can
improve resolution, but govern provides the stronger and more consistent
effect across backbones.

\paragraph{Inform is the main source of reductions in calls and input tokens.}
Relative to the baseline, inform-only reduces model calls from 10.1K to 8.9K
and input tokens from 131.7M to 106.2M with GPT-5 mini. With MiniMax M2.5, it
reduces calls from 30.2K to 21.3K and input tokens from 616.7M to 336.4M. The
same pattern appears when inform is added to govern: relative to govern-only,
the full system reduces calls from 10.8K to 8.9K and from 29.2K to 20.6K,
while input tokens fall from 149.7M to 110.7M and from 589.2M to 289.4M.
Inform therefore substantially reduces executed work both on its own and when
combined with govern.

The corresponding cost reductions also depend on provider-side caching. With
GPT-5 mini, govern-only achieves a higher cache-hit rate than the full system
(80.7\% versus 73.4\%) and is therefore slightly cheaper (\$15.67 versus
\$16.78), despite using more calls and input tokens. Inform-only reaches a
similar cost (\$16.09) through a different mechanism, by reducing the amount
of execution. MiniMax M2.5 already caches 94.7\% of its baseline input, so the
larger savings come from reducing calls and tokens: govern-only lowers cost by
10.2\%, inform-only by 25.1\%, and the full system by 31.8\%.

Overall, the ablation reveals a clear division of roles. Govern contributes
most of the resolution improvement, particularly with MiniMax M2.5, whereas
inform accounts for most of the reduction in calls and input tokens. Their
combination matches the best GPT-5 mini Pass@1, achieves the highest MiniMax
M2.5 Pass@1, and retains substantial efficiency gains on both backbones.

%% file: limitation.tex
\section{Limitations}

First, each configuration is evaluated using a single run per benchmark instance. Because resolution may vary by a few percentage points across independent runs, the reported results are single-run estimates, and we do not characterize run-to-run variance through repeated trials. Although the gains are statistically significant under McNemar's test and resolution and cost improve consistently across all three model-scaffold settings, repeated runs would be needed to fully characterize their stability. Second, we evaluate only on SWE-bench Verified, which draws from Python projects and issue-style bug reports. Whether the same gains hold for other languages, larger repositories, or different task formats is untested. Third, the governing policy is a fixed set of deterministic rules with hand-chosen thresholds. It removes the forms of redundant work those rules describe, can miss patterns they do not cover, and does not adapt its thresholds to the task. Fourth, our controlled evaluation uses a single primary scaffold, with Codex serving only as a transfer test. Behavior on frontier or very small models, and on scaffolds that differ more sharply from the ones we study, remains open.

%% file: conclusion.tex
\section{Conclusion}

In this paper, we presented \tool, a runtime layer that turns a long-horizon coding agent's raw interaction history into explicit execution state: what it has observed, what has changed, and what it has already attempted. \tool maintains this state in a deterministic execution ledger, renders a compact state view into the model input before the agent acts, and evaluates each proposed command against the same state before it executes, all without adding language-model calls or modifying the agent. On all 500 instances of SWE-bench Verified, \tool raises resolution by 8.0 and 5.2 percentage points on two backbones while reducing cost by 28.9\% and 31.8\%, and the same implementation transfers to Codex with only a thin adapter rewritten. Our ablation study shows that the two paths are complementary: governing actions achieves most of the resolution gain, while the state view removes redundant work, mostly by preventing it from being proposed at all. These findings suggest that explicit execution state is an effective and underused lever for long-horizon agents. Context management and execution-state management address distinct costs: the former reduces the representation cost of actions that execute, while the latter can prevent unnecessary actions from occurring. Future work should evaluate their composition and determine whether reductions in action count and per-action representation cost yield cumulative gains.